\documentclass[12pt]{article}

\begin{document}
 
\title {{\bf Problems in Lie Group Theory} \footnote{Submitted to Journ. Opt. B 
(Wigner issue).}}

\author {{ Luis J. Boya } \\
{Departamento de F\'{\i}sica Te\'orica} \\
{Universidad de Zaragoza, E-50009 Zaragoza, Spain } \\
{e-mail: luisjo@posta.unizar.es}}
\date{}

\maketitle 
 
\begin{abstract}

The theory of  Lie groups and representations was 
developed by Lie, Killing, 
Cartan, Weyl and others to a degree of quasi-perfection, in the years
1870-1930 . 
The main topological features of compact simple Lie groups were 
elucidated in
the 40s by H. Hopf, Pontriagin and others.
The exceptional groups were studied by Chevalley, Borel, Freudenthal 
etc. in 1949-1957.
Torsion in the exceptional groups was considered by Toda, Adams etc. 
 in the 80s.
However, one can still ask some questions for which 
the answer is either incomplete or absent, at least to this speaker. 
We would like to raise and discuss some of them in this 
communication.

\end{abstract}

\vfill \eject

\section {Systematic Construction}

The list of compact simple Lie groups contains the classical groups, 
related to $O(n)$, $U(n)$ and $Sp(n)$, and the five exceptional 
structures
$G_{2}$, $F_{4}$, $E_{6}$, $E_{7}$, and $E_{8}$. We shall consider 
the Lie group
and its Lie algebra
simultaneously, referring at times to $G$ and $L(G)$ respectively 
\cite{bib1,bib2}. \\

Given a Lie group in a series $G(n)$ (e.g. the orthogonal, unitary 
\ldots),
how is the group
$G(n+1)$ constructed?  For the {\it orthogonal series} ($B_{n}$ and 
$D_{n}$ in
Cartan\'{}s notation) the answer is simple: given $O(n)$ acting on
itself, that is, the {\it adjoint} (adj) representation, and the 
{\it vector} representation, {\bf n}, it turns out that there is an 
onto map

\begin{equation}
 {\bf n} \wedge {\bf n} \rightarrow adj \ L(O(n))
 \label{eq1}
 \end{equation}
 
 \noindent which satisfies Jacobi identity.  Hence in the direct sum
 $adj + vect$, there is hidden the structure of $L(O(n+1))$
 
 \begin{equation}
     Adj \  O(n) + Vect \  O(n) \rightarrow Adj \  O(n+1)
\label{eq2}
\end{equation}

Dimensions match, but one has to check the Jacobi identity.  In the
simplest case $O(2) \rightarrow O(3)$, if $\{e\}$ describes $O(2)$ and
$\{x,y\}$ the vector representation, defining $[x,y] = e$, the only
thing to check is $[e,x,y] \equiv [e,[x,y]] + cycl.  = 0$, which is
trivial.  By induction, if $L_{ij}$ describes $O(n)$ and $n$ is the 
vectorial,
defining $L_{k,n+1} = x_{k}$, where $x$ is the vectorial $n$, one
checks all Jacobi\'{}s are fulfilled, in a case-by-case procedure. \\

For the {\it unitary series} $SU(n)$ one adds the trivial $U(1)$ plus
the real part of the vector, as $n \times n^{*} = adj + Id$, and the 
balance is

 \begin{equation}
     Adj \ SU(n) + Id + n + n^{*} = Adj \ SU(n+1)
\label{eq3}
\end{equation}

\noindent and checking the Jacobi\'{}s is tedious but it works. \\

For the {\it symplectic series} $Sp(n) = C_{n}$, instead of $U(1)$ 
one adds
$Sp(1)$ to the vector, of complex dimension $2n$

\begin{equation}
     Adj \ Sp(n) + Adj \ Sp(1) + 2(n + n^{*}) = Adj \ Sp(n + 1)
\label{eq4}
\end{equation}

\noindent and again checking Jacobi\'{}s is  messy.  (I would like to
see more clearly why one should add once $Sp(1))$.  \\

For the {\it exceptional groups}, the $F_{4} \  \&  \ E$ series one 
repeats the
orthogonal $(F_{4})$, unitary $(E_{6})$, symplectic $(E_{7})$ and 
orthogonal
$(E_{8})$ method above starting from an orthonormal group and the 
real part
of its Spin representation, to wit (Adams \cite{bib3})

 \begin{equation}
 { Adj \ SO(9) + Spin(9) \ \rightarrow \ Adj \ F_{4}   \  \ \    (36 
+ 16 = 52)}
  \label{eq5}
\end{equation}

 \begin{equation}
     Adj \ SO(10) + Spin(10) + Id \ \rightarrow Adj \ E_{6}  \  \ \ 
(45 + 1 + 2
     \cdot 16 = 78)
\label{eq6}
\end{equation}

 \begin{equation}
     Adj \ SO(12) + Spin(12) + Sp(1) \ \rightarrow Adj \ E_{7} \ \ \ 
(66 + 3 +
     2 \cdot 32 = 133)
\label{eq7}
\end{equation}

  \begin{equation}
  Adj \ SO(16) + Spin(16)  \ \rightarrow Adj \ E_{8} \ \ \  (120 + 
148 = 248)
\label{eq8}
\end{equation}

Notice that $8 + 1$, $8 + 2$, $8 + 4$ and $8 + 8$ appear.  In this
sense the octonions show up as a ``second coming" of the reals,
completed with the spin, not the vector irrep.  One checks that the
antisymmetric product of the spin irrep.  containts the adjoint; for
example for $F_{4}$, $dim(Spin(9) \equiv \ \Delta) = 16$, and
$\Delta \wedge \Delta = 36$ (a $2-form = Adj \ O(9)) + 84$ (a 
3-form). 
This expresses that the $F_{4} \ E_{6-7-8}$ corresponds to the octo,
octo-complex, octo-quater and octo-octo bi-rings, as the
Freudenthal Magic Square confirms \cite{bib4}.  For an explicit
calculation of the Jacobi identity in the last case see \cite{bib5}. 
\\

For $G_{2}$, which has the independent definition as the
automorphism group of octonions, 
we have
the defining sequence $SU(3) \ \rightarrow \ G_{2} \ \rightarrow \ 
S^{6} \
\subset R^{7}$, as $G_{2}$ acts transitive on the 6-sphere of unit
imaginary octonions, and then

\begin{equation}
     Adj \ SU(3) + n + n^{*} \ \rightarrow \ G_{2} \ \ \ (8 + 2 \cdot 
3 = 14)
\label{eq9}
\end{equation}

\noindent whereas the ``conventional" series $SU(3) \ \rightarrow 
SU(4)$,
$dim = 15$, would include a $U(1)$ factor, as above in the unitary
series; this ``unimodular" character of $G_{2}$ is connected, through
exceptional holonomy manifolds, with the compactification in M-theory
$11 \ \rightarrow 4$ \cite{bib6}. \\

So the problem of systematic construction of Lie groups is 
essentially solved.
One would like however to understand better the Jacobi identities, 
and also
why ``other" combinations $adj + rep$ do not lead to new groups 
\ldots \\

The Weyl group, the automorphisms, the center etc.  of these groups
are fairly well understood; Borel \cite{bib7} raises the question, 
for the 
$Spin(4n)$ groups, $n > 2$, whether the two spin irreps $\Delta_{L}$ 
and
$\Delta_{R}$, which are obviously isomorphic, are also isomorphic to 
the vector
irrep: the three have the structure $Spin(4n)/Z_{2}$, (the center of
$Spin(4n)$ is $Z_{2}^{2}$), but only for $n = 2$ is the isomorphism
clear (triality principle), and for $n = 1$ they are clearly not
isomorphism, $Spin(3) \times SO(3) \ne SO(4)$; it is remarkable that
$Spin(4n)$ has no faithful irreps, as any group with more than one
involutive central element.  \\
 
Another, more fundamental question, is the geometry associated to the
exceptional groups, the $E$-series at least.  Are we happy with 
$G_{2}$ as
the automorphism group of the octonions, $F_{4}$ as the isometry of the
octonionic projective plane, $E_{6}$ (in a noncompact form) as the collineations 
of
the same, and $E_{7}$ resp. $E_{8}$ as examples of symplectic resp. 
metasymplectic geometries \cite{bib4,bib8} ?  Many people think this
leaves much to be desired\ldots one would like to understand the 
exceptional
groups at the level we understand the classical groups, as
automorphism groups of some natural geometric objects. \\

A recent paper by Atiyah \cite{bib9} sheds some light in the 
question.  The
first row of the Magic Square \cite{bib8} is just (compact form)

\begin{equation}
    B_{1} : \  RP^{2}=O(3)/O(1) \times O(2) 
    \label{eq10}
    \end{equation}
\begin{equation}
   A_{2} : \  CP^{2}=U(3)/U(2) \times U(1)
  \label{eq11}
    \end{equation}
    
    \begin{equation}
     C_{3} : \  HP^{2} = Sp(3)/Sp(2) \times Sp(1) 
     \label{eq12}
    \end{equation}
    
\begin{equation}
   F_{4} : \   OP^{2}=F_{4}/Spin(9)
    \label{eq13}
    \end{equation}

Now there is a sense in {\it complexifying} the four planes.  The new
projective planes are still homogeneous spaces, giving the second row
of the Magic Square, and a real understanding of $E_{6}$.  We have
the series Group $\rightarrow$ Space $\rightarrow$ Isotropy given by

\begin{equation}
   SU(3) \rightarrow CP^{2} \rightarrow U(2)
    \label{eq14}
    \end{equation}

\begin{equation}
   SU(3) \cdot SU(3) \rightarrow (CP^{2})^{2} \rightarrow (U(2))^{2}
    \label{eq15}
    \end{equation}

\begin{equation}
   SU(6) \rightarrow Gr_{6,2}^{c} \rightarrow S[U(2) \cdot U(4)]
    \label{eq16}
    \end{equation}
    
\begin{equation}
   E_{6} \rightarrow \ X \  \rightarrow Spin(10) \cdot U(1)
    \label{eq17}
    \end{equation}

We lack a clear picture of $X$ \cite{bib9}.
We use only the compact form. The complexified quaternionic plane 
coincides with the grassmaniann of planes in $C^{6}$.  Notice the
naturality of the isotropy groups.  We expect eagerly this analysis to
be extended to the last two rows of the Magic Square.

\section{Topology of Lie groups: Sphere structure}

The gross topology of Lie groups is well-known. The non-compact case 
reduces to compact times an euclidean space (Malcev-Iwasawa theorem). 
The compact case is reduced to a finite factor, a Torus, and a semisimple 
compact Lie group.  H. Hopf determined in 1.941 that the {\it real}
homology of simple compact Lie groups is that of a product of odd
spheres; for example

\begin{equation}
 H_{*}(G_{2};R) = H_{*}(S^{3} \times S^{11};R)
\label{eq22}
\end{equation}

The {\it exponents} of a Lie group are the numbers $i$ such $S^{2i+1}$ is
an allowed sphere; e.g. for $U(n)$, they are $0, 1, \ldots, n-1$.  \\

Can one see this sphere structure directly? The author
has shown \cite{bib10} that in many cases the defining representation
provides a basis for induction, starting from $A_{1} = B_{1} = 
C_{1}$. 
For example, $SU(2) = Sp(1) = Spin(3)$ is exactly $S^{3}$, and as 
$SU(3)$
acts in the defining (vector) irrep in $C^{3} = R^{6} \ \supset \
S^{5}$, the bundle

\begin{equation}
 SU(2) \rightarrow SU(3) \rightarrow S^{5}
\label{eq23}
\end{equation}

\noindent is a principal bundle.  As $\pi_{4}(S^3) = Z_{2}$ classifies
$SU(2)$ bundles over $S^{5}$, and no simple Lie group is a product,
$SU(3)$ is the {\it unique} non-trivial bundle over $S^{5}$ with fiber
$SU(2)$, in the sense that the square is trivial; hence we dare write

\begin{equation}
  SU(3) = S^{3} \  ( \times  \ S^{5}
\label{eq24}
\end{equation}

\noindent as a finite twisted product.
For the unitary and symplectic series the method works perfectly (see
\cite{bib10}); indeed, this is connected with the fact that neither the
U-series nor the Sp-series have torsion \cite{bib7}.  The exponents are
succesive in $U(n)$ and jump by two in $Sp(n)$. \\

But for the orthogonal series one has to consider 
some other manifolds besides  spheres, with the same real homology; 
is this an imperfection?  No!  It accounts for the {\it torsion}.  The
place of spheres is played by some Stiefel manifold, and this
introduces (precisely) 2-torsion: in fact, $Spin(n)$, $n \geq 7$ and
$SO(n)$, $n \geq 3$, have 2-torsion.  The low cases $Spin(3,4,5,6)$
coincide with $Sp(1)$, $Sp(1) \times Sp(1)$, $Sp(2)$ and $SU(4)$, and have no
torsion.  \\

For the exceptional groups, let us start with the smallest, $G_{2}$; the
structure diagram is \cite{bib11} 

\begin{equation} \begin{array}{ccccc}
SU(2) & = & SU(2) & & \\
\downarrow &              & \downarrow &              &  \\ 
SU(3)      & \rightarrow  & G_{2}      &  \rightarrow & S^{6}  \\
\downarrow &              & \downarrow &              & \parallel \\
S^{5}      & \rightarrow  & M_{11}     & \rightarrow  & S^{6}

\label{eq45}
\end{array} \end{equation}

\noindent where $M_{11}$ is again a Stiefel manifold, with real homology
like $S^{11}$, but with 2-torsion.  Hence 

\begin{equation}
 H_{*}(G_{2};R) = H_{*}(S^{3} \times S^{11};R),
\label{eq25}
\end{equation}

\noindent the correct result.  For $F_{4}$ we do not get the sphere
structure from any irrep, and in fact $F_{4}$ has 2- and 3-torsion. 
Does the 3-torsion of $F_{4}$ come from the Euler Triplet, i.e. Euler number of 
$(F_{4}/Spin(9) =$ Moufang  Plane \footnote{ The projective plane 
over octonions was discovered by Ruth Moufang in 1.932. It makes little sense
to call it the Cayley plane;
A. Cayley was not even the first discoverer of the octonions!}) $= 3$?  \cite{bib12}.  \\

There is no torsion in the U- and Sp-series, 
2-torsion only in $Spin(n)$ and $G_{2}$, as referred to above.  
Now 2- and 3-Torsion appears in $F_{4}$ (as mentioned), $E_{6}$ and
$E_{7}$. We have here no comment to offer; in particular, there is no clue
that we see for torsion on the twisted sphere product and the natural
actions of these groups, nor for $E_{8}$.  But\ldots  \\

$E_{8}$ has 2-, 3- and 5-torsion \cite{bib2}!  Where on earth the 5-torsion
comes from?  I should pinpoint two hints: (i): The Coxeter number
$(dim-rank)/rank$ of $E_{8}$ is $30=2 \cdot 3 \cdot 5$, in fact a mnemonic for the
exponents of $E_{8}$ is: they are the coprimes up to 30, namely (1, 7, 11,
13, 17, 19, 23, 29) ; (ii) The first perfect numbers are 6, 28 and 496,
associated to the primes 2, 3 and 5 (these Mersenne numbers are $2^{p-1} \cdot
(2^{p} - 1)$, $p$ and $2^{p} - 1$ primes).  And the reader will recall that $496 = dim \
O(32) = dim \  E(8) \times E(8)$.  Why the square?  It happens also in $O(4)$, dim
= 6 (prime 2), as $O(4) \sim  O(3) \times O(3)$; even $O(8)$ (prime 3) is
like $S^{7} \times S^{7} \times G_{2}$. \\

These are not real problems, but features for which we should expect a
better explanation. 

\section{Other topological features}

\subsection{Capicua}

The sphere structure of compact simple Lie
groups has a curious ``capicua" \footnote{This catalan word ({\it cap i cua}
= head and tail) is much more expressive that the greek {\it palindrome}. 
Capicua RNA sequences are very important in the RNA world which
presumably ruled on earth before organized life.} form: the exponents
are symmetric from each end; for example, for $E_{6}$ and $E_{7}$:

\begin{equation}
 exponents \  of \  E_{6}:1,4,5,7,8,11. \   Differences:3,1,2,1,3
\label{eq26}
\end{equation}

\begin{equation}
 exponents \  of \  E_{7}:1,5,7,9,11,13,17. \   Differences:4,2,2,2,2,4
\label{eq27}
\end{equation}

This question was raised by Chevalley \cite{bib13}, and still (I) do not
understand it.

\subsection{Supersymmetry}

The real homology algebra of  a simple Lie group is a Grassmann algebra, as it is 
generated by odd (i.e., anticonmutative) elements. However, from them we can get, 
in the enveloping algebra, multilinear symmetric forms, one for each generator; 
the construction is standard \cite{bib13}; in physics they are called
the Casimir invariants, in mathematics the invariants of the Weyl group. \\

For example, for $SU(3)$, we have the quadratic and the cubic
invariant \cite{bib14}

\begin{equation}
 I_{2}(x,y) =  Tr \ ad \ x \ ad \ y,, \ \ \ I_{3}(x,y,z) =\{x,y \vee z\}
\label{eq28}
\end{equation}

Is this a fact of life, or an indication of a hidden odd-even symmetry 
(supersymmetry)?  It was remarked already in \cite{bib10}. \\

There is probably a more profound relation with 
supersymmetry, which we are just starting to notice; it was discovered by P. Ramond 
\cite{bib15}, and the mathemathical basis is been clarified by B. Kostant
\cite{bib16}.  The first example is with the pair $F_{4}-B_{4}$: it turns out
that $F_{4}/B_{4}$ is the Moufang plane
$OP^{2}$ and 

\begin{equation}
 dim \ Weyl(F_{4})/  Weyl(B_{4}) = 1152/384 = Euler \ (OP^{2}) = 3
\label{eq29}
\end{equation}

\noindent as $b_{0}=b_{8}=b_{16}=1$, others $b\acute{}s = 0$ in $OP^{2}$.
Kostant now says: any irrep of $F_{4}$ generates three of $B_{4}$, in many
cases supersymmetric, that is, the dimensions of spinors (faithful irrep  of
$Spin(9)$) and tensors (faithful of $SO(9)=Spin(9)/Z_{2}$) match; for example
(negative signs for spinors) 

\begin{equation}
 The \  identical \ irrep \ of F_{4} \  generates \ the \ triplet: \ +44-128+84 \ of \ Spin(9)
\label{eq30}
\end{equation}

\noindent corresponding, in physics, to the 11-dimensional maximal
Supergravity multiplet with graviton, gravitino and the 3-form C,
which in $M$-theory is radiated by the M2 brane.  Ramond found many
triplets with fermi/bose ( = spinorial vs.  vectorial irreps)
matching, and also many matchings in the dimensions of the Casimir
invariants (but not perfect!). \\

\noindent 
We think this is a very important development; it points out to
explain, for the first time, the existence of supersymmetry in
physics!  The theory \cite{bib16} is that for any pair $H \ \subset \ G$ of
Lie groups, $G$ semisimple and $H$ reductive subgroup (i.e. it can contain
$U(1)$ subgroups), the $Pin(2N)$ irrep, $2N = dim \ G - dim \ H$, splits in
$\chi$ basic irreps of $H$, where 

\begin{equation}
 \chi = Euler (G/H) = dim [Weyl(G)/Weyl(H)]
\label{eq31}
\end{equation}

The analogous analysis for $CP^{2} = SU(3)/U(2)$ 
and $HP^{2}= Sp(3)/[Sp(1) \times Sp(2)]$, also with Euler number $=3$,
corresponds to the supersymmetric hypermultiplet and Yang-Mills
multiplet, respectively \cite{bib15}.  That the 3 simply connected projective
planes display the standard Susy, scalar in 6D, vector in 10D and
tensor in 11D is thrilling!  \\

We explain the procedure for the identity 
irrep of $SU(n+1)$; for instance let us  
consider in $CP^{n} = SU(n+1)/U(n)$ the diagram 

\begin{equation} \begin{array}{ccccc}
 &  & Z_{2} & = & Z_{2} \\
 & & \downarrow & & \downarrow  \\
U(n) & \rightarrow & Spin(2n) & \rightarrow & X \\
\parallel  & & \downarrow & & \downarrow \\
U(n) & \rightarrow & SO(2n)  & \rightarrow & X/2
\label{eq55}
\end{array} \end{equation}

\noindent where the nature of $X$ needs not concern us (it is $CP^{3}$
for $n = 3$).  The irrep of $Pin(2n)$ of dim $2^{n}$ splits, regarding $U(n)$, in
all the antisymmetric forms (p-forms); for example for $n=4$ 

\begin{equation}
 \Delta_{L} + \Delta_{R} = [0] + [1] + [1^{2}] +[1^{3}] + [1^{4}]
\label{eq32}
\end{equation}

\begin{equation}
 2^{4} = 16 = 1^{+} + 4^{-} + 6^{+} + 4^{-} + 1^{+}
\label{e33}
\end{equation}

The $U(1)$ factor in $U(n) = [ SU(n) \times U(1)]/Z_{n}$ gives the 
sign (grading) which amounts to a generalization of supersymmetry;
notice the number of summands, 5, is Euler($CP^{4}$).  Kostant character
formula \cite{bib16} ammounts to substracting, instead of adding, the two spin
irreps, and tensoring by any irrep of $SU(n + 1)$: the alternating multiplet
in the r.h.s, which makes sense in the Grothendieck representation
ring $R(G)$, exhibits a generalized supersymmetry.  For extensions to
the symplectic and octonionic cases see \cite{bib17}.

\section{Representations}

For {\it finite} groups there is a {\it duality}, a kind of Fourier equivalence, 
between conjugacy classes and irreducible representations in the
following sense (Frobenius): the number is the same, and the irreps
are obtained from the central idempotents in which the classes, 
as centrals in the group algebra, are decomposed spectrally. 
In particular the group algebra splits in sum of matrix algebras, 
one per class, containing the irreps as many times as the dimension: 
Burnside\'{}s formula reflects this \cite{bib18}: 

\begin{equation}
 ord \  G = \sum \ d_{i}^{2}
\label{eq34}
\end{equation}

\noindent where $i$ runs through the irreps $\leftrightarrow$ classes. \\

The question we want to pose now is this: how does this correspondence from 
conjugacy classes vs.  irreps generalize for compact Lie groups?
For compact {\it abelian} Lie groups $A$ it is very clear: it is Pontriagin
duality, trading the circle by the integers as many times as $rk A$. \\

It should be possible to explain the gross features of the discrete 
lattice of irreps of a compact simple Lie group by the geometry of the 
compact manifold (or rather, orbifold) of the conjugacy classes. How? \\

I can only recall the simplest case, $SU(2)$; the set of classes is a
segment (labelled by the rotation angle), and the irreps are the
nonnegative integers.  For a group of rank $r$, the set of classes is a
compact ``manifold" of dimension $r$, and consequently the lattice of irreps
is labeled by $r$ integers.  How does one go beyond this?  One would
like to do Fourier analysis in the center (class functions) of the
$L^{1}(G)$ convolution algebra of functions, and distillate the lattice of
irreps\ldots  \\

The irreps of the classical and exceptional structures have a 
set of order $r$ again called  {\it primitive} irreps, in the sense that the 
ring $R(G)$ is generated by them; they correspond one to one to the nodes of 
the Dynkin diagram. Our next question is: How are these irreps selected? \\

For the A-, B-, C- and D-series, 
it is fairly clear: they are the natural (or defining) irrep, 
plus some p-forms, traceless in the C-series, plus the spin(s) irreps. 
in the O-series. Indeed the different root in the symplectic series $C_{n}$ is
the irrep $[1^{n}] - [1^{n-2}]$ with (complex) singularized dimension
$2(2n+1)(2n)\ldots (n + 3)/n!$.  \\

Now for the primitive irreps of the exceptional groups \cite{bib20}:
for $G_{2}$ are the natural irrep (7) and the adjoint (14).  For $F_{4}$, the
natural (3x3 octonionic hermitean traceless, 26) and the adjoint (52). 
The other two are: $273 = 26 \cdot 25/2 - 52$ and $1274 = 52 \cdot 51/2 -
52 = (26 \cdot 25 \cdot 24/3! - 52)/2$, natural constructs in both cases.  \\

For $E_{6}$ the natural is $27$ (given by the $3 \times 3$ hermitean
complex-octonionic Jordan algebra), and the adjoint is 78.  As the
center is $Z_{3}$, there are complex irreps \cite{bib19} (this is why
$E_{6}$ is a candidate for Grand Unified Theories, as it can
accomodate chiral multiplets).  Besides $27^{*}$, $351 = 27 \cdot
26/2$, $351^{*}$ and the real $2925 =27 \cdot 26/2 \cdot 3$, exterior
products.  Also $2925 = 78 \cdot 77/2 - 78$.  \\

For $E_{7}$, 56 defining and 133 adjoint; then $1539=56 \cdot 57/2-1$, $27664
=56 \cdot 55 \cdot 54/2 \cdot 3 - 56$;  $365750=56 \cdot 55 \cdot 54
\cdot 53/4!  - 56 \cdot 55/2$ all correspond to antisymmetric traceless
products, as $E_{7}$ is symplectic.  $8645=133\cdot 132/2 - 133$ is also
natural.  There is still the 912, for which T. Smith (Atlanta, GA) (personal
communication) proposes $912 = 16 \times 56 + 16 \times 1$.

Finally, for $E_{8}$, $248$ is both the natural and the 
adjoint (unique case among Lie groups). All except $3875$ and $147250$ are obtained,
again, from traceless p-forms on $248$ \cite{bib21}.

\section*{Dedicatory}

The man who made group theory and representations accesible and 
useful for physics was Eugene P. Wigner, whose 100th birthday we 
gather 
here to celebrate. I offer this modest contribution to his memory.

\section*{Acknowledgements}

I have talked many of the topics presented here with M. Santander
(Valladolid), who was very helpful.  Also discussions with A. Segui
(Zaragoza) and J. Mateos (Salamanca) are gratefully acknowledged.

\end{document}